\documentclass[%
reprint,
superscriptaddress,
showpacs,
 amsmath,amssymb,
 aps,
 prl,
]{revtex4-1}

\usepackage{graphicx}
\usepackage{dcolumn}
\usepackage{bm}
\usepackage{url}
\usepackage{soul}

\usepackage[d]{esvect}
\usepackage{calc}
\usepackage{color}

\begin{document}

\preprint{APS/123-QED}

\title{
Measurement of the transverse single-spin asymmetry \\ 
in $ p^\uparrow+p \rightarrow W^{\pm}/Z^0$ at RHIC}




\affiliation{AGH University of Science and Technology, FPACS, Cracow 30-059, Poland}
\affiliation{Argonne National Laboratory, Argonne, Illinois 60439}
\affiliation{Brookhaven National Laboratory, Upton, New York 11973}
\affiliation{University of California, Berkeley, California 94720}
\affiliation{University of California, Davis, California 95616}
\affiliation{University of California, Los Angeles, California 90095}
\affiliation{Central China Normal University, Wuhan, Hubei 430079}
\affiliation{University of Illinois at Chicago, Chicago, Illinois 60607}
\affiliation{Creighton University, Omaha, Nebraska 68178}
\affiliation{Czech Technical University in Prague, FNSPE, Prague, 115 19, Czech Republic}
\affiliation{Nuclear Physics Institute AS CR, 250 68 Prague, Czech Republic}
\affiliation{Frankfurt Institute for Advanced Studies FIAS, Frankfurt 60438, Germany}
\affiliation{Institute of Physics, Bhubaneswar 751005, India}
\affiliation{Indian Institute of Technology, Mumbai 400076, India}
\affiliation{Indiana University, Bloomington, Indiana 47408}
\affiliation{Alikhanov Institute for Theoretical and Experimental Physics, Moscow 117218, Russia}
\affiliation{University of Jammu, Jammu 180001, India}
\affiliation{Joint Institute for Nuclear Research, Dubna, 141 980, Russia}
\affiliation{Kent State University, Kent, Ohio 44242}
\affiliation{University of Kentucky, Lexington, Kentucky, 40506-0055}
\affiliation{Korea Institute of Science and Technology Information, Daejeon 305-701, Korea}
\affiliation{Institute of Modern Physics, Chinese Academy of Sciences, Lanzhou, Gansu 730000}
\affiliation{Lawrence Berkeley National Laboratory, Berkeley, California 94720}
\affiliation{Max-Planck-Institut fur Physik, Munich 80805, Germany}
\affiliation{Michigan State University, East Lansing, Michigan 48824}
\affiliation{National Research Nuclear Univeristy MEPhI, Moscow 115409, Russia}
\affiliation{National Institute of Science Education and Research, Bhubaneswar 751005, India}
\affiliation{National Cheng Kung University, Tainan 70101 }
\affiliation{Ohio State University, Columbus, Ohio 43210}
\affiliation{Institute of Nuclear Physics PAN, Cracow 31-342, Poland}
\affiliation{Panjab University, Chandigarh 160014, India}
\affiliation{Pennsylvania State University, University Park, Pennsylvania 16802}
\affiliation{Institute of High Energy Physics, Protvino 142281, Russia}
\affiliation{Purdue University, West Lafayette, Indiana 47907}
\affiliation{Pusan National University, Pusan 46241, Korea}
\affiliation{University of Rajasthan, Jaipur 302004, India}
\affiliation{Rice University, Houston, Texas 77251}
\affiliation{University of Science and Technology of China, Hefei, Anhui 230026}
\affiliation{Shandong University, Jinan, Shandong 250100}
\affiliation{Shanghai Institute of Applied Physics, Chinese Academy of Sciences, Shanghai 201800}
\affiliation{State University Of New York, Stony Brook, NY 11794}
\affiliation{Temple University, Philadelphia, Pennsylvania 19122}
\affiliation{Texas A\&M University, College Station, Texas 77843}
\affiliation{University of Texas, Austin, Texas 78712}
\affiliation{University of Houston, Houston, Texas 77204}
\affiliation{Tsinghua University, Beijing 100084}
\affiliation{United States Naval Academy, Annapolis, Maryland, 21402}
\affiliation{Valparaiso University, Valparaiso, Indiana 46383}
\affiliation{Variable Energy Cyclotron Centre, Kolkata 700064, India}
\affiliation{Warsaw University of Technology, Warsaw 00-661, Poland}
\affiliation{Wayne State University, Detroit, Michigan 48201}
\affiliation{World Laboratory for Cosmology and Particle Physics (WLCAPP), Cairo 11571, Egypt}
\affiliation{Yale University, New Haven, Connecticut 06520}

\author{L.~Adamczyk}\affiliation{AGH University of Science and Technology, FPACS, Cracow 30-059, Poland}
\author{J.~K.~Adkins}\affiliation{University of Kentucky, Lexington, Kentucky, 40506-0055}
\author{G.~Agakishiev}\affiliation{Joint Institute for Nuclear Research, Dubna, 141 980, Russia}
\author{M.~M.~Aggarwal}\affiliation{Panjab University, Chandigarh 160014, India}
\author{Z.~Ahammed}\affiliation{Variable Energy Cyclotron Centre, Kolkata 700064, India}
\author{I.~Alekseev}\affiliation{Alikhanov Institute for Theoretical and Experimental Physics, Moscow 117218, Russia}
\author{A.~Aparin}\affiliation{Joint Institute for Nuclear Research, Dubna, 141 980, Russia}
\author{D.~Arkhipkin}\affiliation{Brookhaven National Laboratory, Upton, New York 11973}
\author{E.~C.~Aschenauer}\affiliation{Brookhaven National Laboratory, Upton, New York 11973}
\author{A.~Attri}\affiliation{Panjab University, Chandigarh 160014, India}
% ...

\collaboration{STAR Collaboration}\noaffiliation

\date{\today}

\begin{abstract}
We present the measurement of the transverse single-spin asymmetry of weak boson production in transversely polarized proton-proton collisions at $\sqrt{s} = 500~\text{GeV}$ by the STAR experiment at RHIC. 
The measured observable is sensitive to the Sivers function, one of the transverse momentum dependent parton distribution functions, which is predicted to have the opposite sign in proton-proton collisions from that observed in deep inelastic lepton-proton scattering. These data provide the first experimental investigation
of the non-universality of the Sivers function, fundamental to our understanding of QCD. 
\end{abstract}

\pacs{24.85.+p, 13.38.Be, 13.38.Dg, 14.20.Dh}
\maketitle


%

During the past decade there have been tremendous efforts towards understanding the three-dimensional partonic structure of the proton.  
One way to describe the 2+1 dimensional structure of the proton in momentum space is via transverse-momentum-dependent parton distribution functions (TMDs)~\cite{Aybat:2011}, 
which encode a dependence on the intrinsic transverse momentum of the parton $k_T$ in addition to the longitudinal momentum fraction $x$ of the parent proton carried by the parton. There are eight TMDs that are allowed by parity invariance~\cite{Meissner:2009}. Of particular interest is the Sivers function~\cite{Sivers:1991}, $f^{\perp}_{1T}$, which describes the correlation between the intrinsic transverse momentum of a parton and the spin of the parent proton. It may be described as the parton density of the vector structure
 $(\vec{P} \times \vec{k}_T)\cdot \vec{S}$, where $\vec{P}$ and $\vec{S}$ are respectively the proton momentum and spin vectors. 
In $p+p$ collisions in which one of the proton beams is transversely polarized, the Sivers function can be accessed through measurements of the transverse single-spin asymmetry (TSSA) in Drell-Yan (DY)   or $W^{\pm}/Z^{0}$  boson production, which is defined as: 
$(\sigma_\uparrow - \sigma_\downarrow) / (\sigma_\uparrow + \sigma_\downarrow)$,  
where $\sigma_{\uparrow (\downarrow)}$ is the cross section measured with the spin direction of the proton beam pointing up (down).

In addition to providing access to the three-dimensional structure of the nucleon, there are non-trivial predictions for the process dependence of the Sivers function stemming from gauge invariance. 
In semi-inclusive deep inelastic scattering (SIDIS), the Sivers function is associated with a final-state effect through gluon exchange between the struck parton and the target nucleon remnants~\cite{DIS_Sivers}. 
In $p+p$ collisions, on the other hand, the Sivers asymmetry originates from the initial state of the interaction for the DY process and $W^{\pm}/Z^{0}$ boson production. 
As a consequence, the gauge invariant definition of the Sivers function 
predicts the opposite sign for the Sivers function in SIDIS compared to processes with color charges in the initial state and a colorless final state, such as $p+p \rightarrow \text{DY}/W^{\pm}/Z^{0}$~\cite{Collins:2002}

\begin{equation}
f^\text{SIDIS}_{q/h^\uparrow} (x, k_T, Q^{2}) = - f^{p+p \rightarrow \text{DY}/W^{\pm}/Z^{0}}_{q/h^\uparrow} (x, k_T, Q^{2}).
\end{equation}
This non-universality of the Sivers function is a fundamental prediction from the gauge invariance of the theory of Quantum Chromo-Dynamics (QCD) and is based on the QCD factorization formalism~\cite{Collins:2002,factorization}. 
The experimental test of this sign change is a crucial measurement in hadronic physics~\cite{NSAC}, and will provide an important test of our understanding of QCD factorization.

DY and $W^{\pm}$, $Z^{0}$ production in $p+p$ collisions provide the two scales required to apply the TMD framework to transverse single-spin asymmetries. A hard scale is given by the photon virtuality ($Q^{2}$) or by the mass of the produced boson ($M^{2} \sim Q^{2}$), while a soft scale of the order of the intrinsic $k_T$ is given by the transverse momentum.  
While a measurement of the TSSA in Drell-Yan production at 
forward pseudorapidities ($\eta >  3$) is experimentally very challenging, requiring severe background suppression and substantial integrated luminosity, a TSSA measurement in weak boson production offers several unique advantages. 
Due to the high $Q^2 \simeq M^2_{W/Z}$  scale provided by the large boson mass ($M_{W/Z}$), 
the measurement of the TSSA amplitude ($A_N$) in weak boson production provides a stringent test 
of the evolution of the TMDs~\cite{Kang:2014}, which like for other asymmetries are expected to partially cancel in the ratio of polarised and unpolarised cross section.
The rapidity dependence of $A_N$ for the $W^{+}(W^{-})$ boson, which is produced through $u+\bar{d}~(d +\bar{u})$ fusion, provides an essential input to reduce the uncertainty on the Sivers function for light sea quarks. 
That Sivers function, determined by fits to SIDIS data~\cite{Kang:2014} in a Bjorken-$x$ range where the asymmetry of the $\bar{u}$ and $\bar{d}$ unpolarized sea quark densities~\cite{E866} can only be explained by strong non-perturbative QCD contributions, is essentially unconstrained. 


The $A_N$ of the lepton produced in $W^{\pm}$ decay is predicted~\cite{Metz:2011,Kang:2009bp} to vary rapidly with the lepton kinematics, having a non-zero value in only a narrow region in lepton transverse momentum and pseudorapidity. 
On the other hand, the asymmetry is predicted to have a sizable value over a large range of the produced boson kinematics~\cite{Kang:2009bp}, its actual magnitude depending on the TMD evolution~\cite{Kang:2014}. Therefore, in measuring $A_N$, it is preferable to fully reconstruct the $W$ boson. 


In this letter, we report the measurement of $A_N$ for weak bosons in 
proton-proton collisions at $\sqrt{s}=500$~GeV with transversely polarized beams by the STAR experiment at RHIC. The data sample used in this analysis was collected in the year 2011, and corresponds to a recorded integrated luminosity of 25 pb$^{-1}$. 
The beam polarization was measured using Coulomb-nuclear interference proton-carbon polarimeters,
calibrated with a polarized hydrogen gas-jet target. 
The average beam polarization for the data set used in the present analysis was 53\%, with a relative scale uncertainty of $\Delta P/P=$~3.4\%~\cite{RHIC_polarimetry}.
The subsystems of the STAR detector~\cite{STAR_detector} used in this measurement are the Time Projection Chamber (TPC)~\cite{TPC_detector}, providing charged particle tracking for pseudorapidity
$|\eta| \le 1.3$, and the Barrel Electromagnetic Calorimeter (BEMC)~\cite{BEMC_detector}, covering the full azimuthal angle $\phi$ for $|\eta| < 1$.

In this analysis, data were recorded using a calorimeter trigger requirement of 12 GeV of transverse energy $E_{T}$ in a $\Delta \eta \times \Delta \phi$ region of $\sim0.1 \times 0.1$ of the BEMC.
Based on STAR previous analyses of weak boson longitudinal spin asymmetries~\cite{STAR:Wlong} and cross sections~\cite{STAR:Wxsec}, we selected a data sample characterized by the $W\rightarrow e \nu$ signature, requiring an isolated electron with $P_{T}^{e} > 25$~GeV/c within the BEMC acceptance ($|\eta|<1$). 
In reconstructing the momentum of the decay electron, its energy was measured in the BEMC and its trajectory using the TPC.    

To ensure the isolation of the decay electron, it is required that the ratio of the sum of the electron momentum and energy, $(P^{e}+E^{e})$, over the sum of the momenta and energies of all the particles contained in a cone with a radius $R=\sqrt{(\eta^2+\phi^2)}=0.7$ around the decay electron track, $\Sigma_{Rcone=0.7}[P^{\textrm{tracks}}+E^{\textrm{cluster}}]$, must be larger than 0.9. 
All tracks must come from a single vertex with $|Z_{\textrm{vertex}}| < 100$~cm.

We define the variable $P_{T}$-balance, $\vec{P}_{T}^{bal}$, as the vector sum of the decay electron candidate $\vec{P}_{T}^{e}$ and the transverse momentum of the hadronic recoil, $\vec{P}_{T}^{\textrm{recoil}}$. 
The latter is calculated as the vector sum of  the transverse momenta of all tracks with $P_{T}>200$~MeV/c, excluding the decay electron candidate, and the $E_{T}$ of all 
clusters in the BEMC without a matching track and with an energy above the noise threshold of 200~MeV.
In order to reject QCD background events the scalar variable $(\vec{P}_{T}^{bal} \cdot \vec{P}_{T}^{e}) / |\vec{P}_{T}^{e}|$ is required to be larger than 18~GeV/c. 

After applying all the selection criteria, the remaining electron candidates are sorted by charge. Charge misidentification was minimized by requiring that the lepton transverse momentum, $P_T^{e-track}$, as measured by the TPC track, satisfies the condition $0.4 < E_T^e/P_T^{e-track} < 1.8$ for both charge signs.  The contamination from incorrectly assigned events is estimated to be ~ 0.004\%.
The selection yields final data samples of 1016 $W^{+}$ events and 275 $W^{-}$ events for $0.5~\text{GeV}/c < P_T^W < 10~\text{GeV}/c$. 

In this work, the $W^{\pm}$ kinematics were, for the first time, fully reconstructed for a spin observable, following the analysis techniques as previously used at the Tevatron and LHC experiments, e.g., see~\cite{Tevatron-LHC-Wrecon}. 
In reconstructing the boson kinematics, the momentum of the neutrino produced in the leptonically decayed $W \rightarrow l + \nu$
can only be deduced indirectly from transverse momentum conservation: $\vec{P}^{W}_{T} = - \vec{P}_{T}^{\textrm{recoil}}$. 
At the STAR detector, due to its limited pseudorapidity acceptance,
the challenge with measuring the momentum from the hadronic recoil
is that particles at high pseudorapidities are not detected. However, particles at high pseudorapidity typically carry only a small fraction of the total transverse momentum. 
The unmeasured tracks and clusters are accounted for by using an event-by-event Monte Carlo (MC) correction to the data. The correction factor $c_{i}$ to the measured 
$W$ transverse momentum in the $i$-th  bin is defined as

\begin{equation}
c_{i}=\frac{P^{W}_{T,i}(\textrm{true})}{P^{\textrm{recoil}}_{T,i}(\textrm{reconstructed})},
\end{equation}
where $P^{W}_{T,i}(\textrm{true})$ is the $P_{T}$ of the $W$ generated by the MC and $P^{\textrm{recoil}}_{T,i}(\textrm{reconstructed})$ is the $P_{T}$ of the recoil reconstructed in each $i$-th bin after a full simulation of the detector and applying all the selection requirements. 
For each event, the measured value of the boson $P_{T}$ was corrected by randomly sampling a value from the corresponding $P_{T}$-bin of the normalized correction factor distribution. 

\begin{figure}[htbp]
\begin{center}
\includegraphics[scale=0.45]{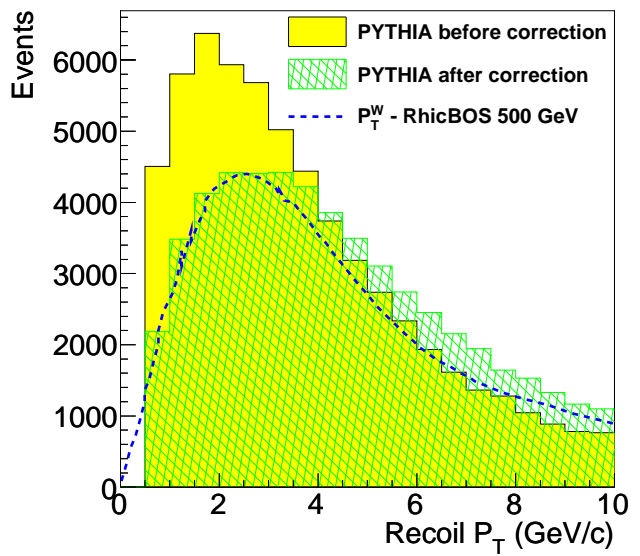}
\end{center}
\caption{[{\it Color online}] $P_{T}^{\textrm{recoil}}$ distribution of events simulated with PYTHIA 6.4 and reconstructed before ({\it yellow}) and after ({\it hatched green}) the $P_{T}$ correction has been applied, is compared with predictions from RhicBOS ({\it dashed blue}).} 
\label{fig:plot_PtCorr} 
\end{figure}

In identifying the hadronic recoil from the tracks and clusters,  
events are rejected if the total $P_{T}^{\textrm{recoil}} <0.5$~GeV/c, 
a region where the correction factor becomes large and has a broad distribution.
The MC simulation using PYTHIA 6.4~\cite{PYTHIA6.4} with the ``Perugia 0'' tune~\cite{PerugiaTune} shows that after the correction has been applied, the reconstructed $P_{T}$ of the $W$ boson agrees with the independent prediction from RhicBOS~\cite{RhicBOS}, as shown in Fig.~\ref{fig:plot_PtCorr}. The MC samples have been passed through the GEANT 3~\cite{GEANT3} simulation of the STAR detector and are embedded into 
events from a zero-bias trigger.

Knowing its transverse momentum, the longitudinal component of the neutrino's momentum, $P_{L}^{\nu}$, can be reconstructed, solving the quadratic equation for the invariant mass of the produced boson

\begin{equation}
\label{eq_InvMass}
M^{2}_{W}=(E_{e}+E_{\nu})^{2} - (\vv P_{e}+\vv P_{\nu})^{2},
\end{equation}
where the nominal value of the $W$ mass is assumed.  
Equation~\ref{eq_InvMass} leads to two possible solutions for $P_{L}^{\nu}$. A MC study showed that for $|P_{L}^{\nu}| < 50$~GeV/c, corresponding to a $W$ boson rapidity $|y^{W}| < 0.6$, the solution with the smaller absolute value gives on average a more accurate reconstruction of the originally generated $W$ boson kinematics. 

\begin{figure}[htbp]
\centering
\includegraphics[scale=0.46]{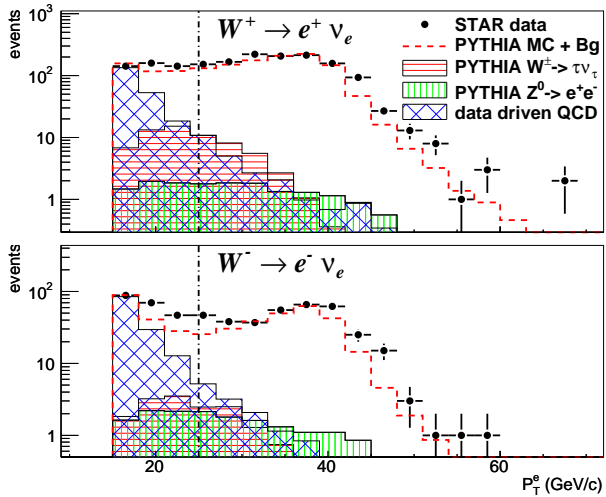}
\caption{[{\it Color online}] Estimated background contributions are 
shown for the $W^{+}$ ({\it upper}) and the $W^{-}$ ({\it lower}) data samples respectively. Vertical line marks the minimum $P^{e}_{T}$ value in this analysis.}  
\label{fig:plot_WZ_backgrounds} 
\end{figure}

\begin{table}[htbp]
\centering
\begin{tabular}{| l | c | c  |  c |}
\hline
Process & $W^{\mp}\rightarrow \tau^{\mp} \bar{\nu}_{\tau}$ & $Z^{0} \rightarrow e^{+}e^{-}$ & QCD \\
\hline
$W^{+}$ (B/S) & $1.89\%\pm0.04\%$ &  $0.79\%\pm0.03\%$ & $1.6\%\pm0.09\%$ \\
$W^{-}$ (B/S) & $1.77\%\pm0.10\%$ & $2.67\%\pm0.10\%$ &   $3.39\%\pm0.23\%$ \\
\hline
\end{tabular}
\caption{Background over signal in the $W^{+}$ and $W^{-}$ samples respectively.}
\label{Tab:BoS}
\end{table}

\begin{figure*}[htbp]
  \centering
  \includegraphics[width=0.8\textwidth]{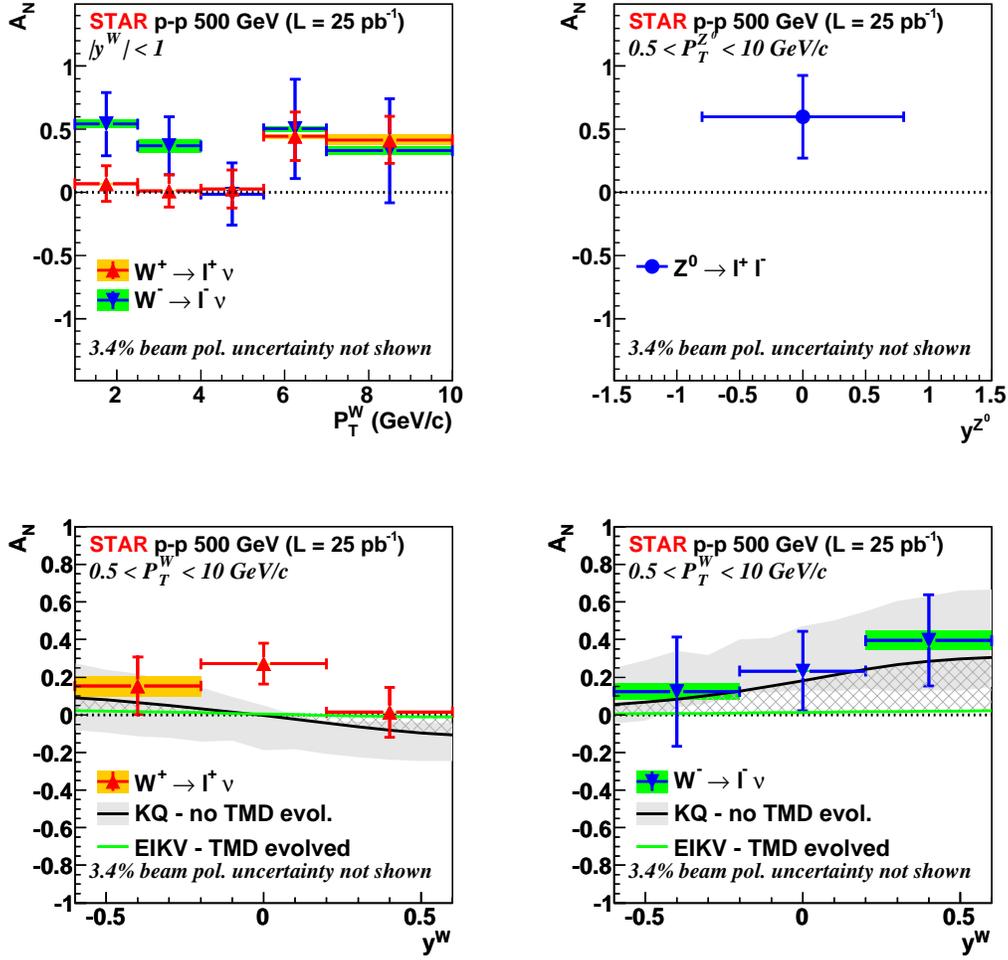}
  \caption{[{\it Color online}] The amplitude of the transverse single-spin asymmetry for $W^{\pm}$ and $Z^{0}$ boson production measured by STAR in proton-proton collisions at $\sqrt{s}=500$~GeV with a recorded luminosity of 25~$\text{pb}^{-1}$. The solid gray bands represent the uncertainty on the KQ~\cite{Kang:2009bp} model due to the unknown sea quark Sivers function. The crosshatched region indicates the current uncertainty in the theoretical predictions due to TMD evolution.}
  \label{Fig:W-An}
\end{figure*}

\begin{figure*}[htb]
  \centering
  \includegraphics[width=0.79\textwidth]{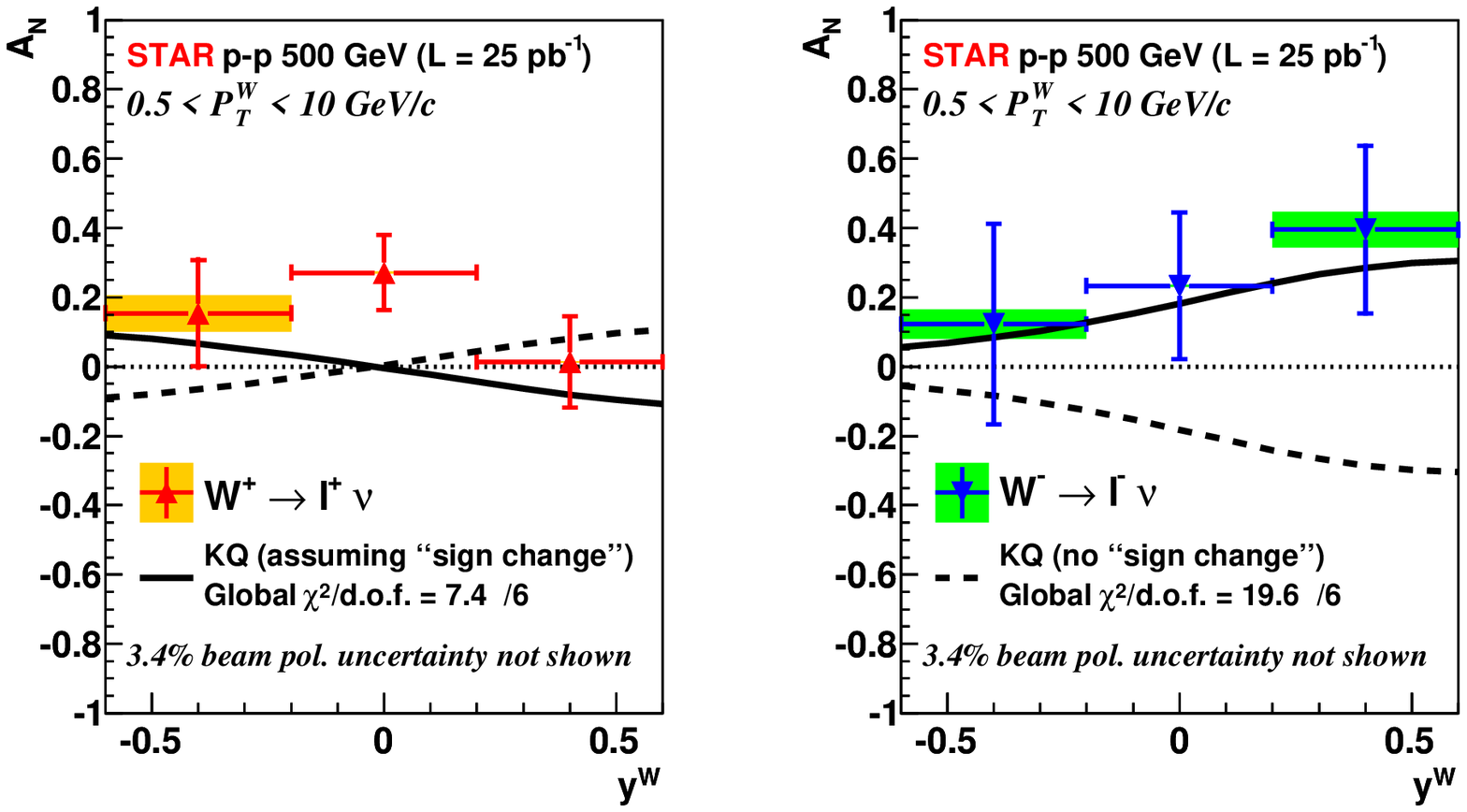}
  \caption{[{\it Color online}] Transverse single-spin asymmetry amplitude for $W^{+}$ ({\it left plot}) and $W^{-}$ ({\it right plot}) versus $y^{W}$ compared with the non TMD-evolved KQ~\cite{Kang:2009bp} model, assuming ({\it solid line}) or excluding ({\it dashed line}) a sign change in the Sivers function.}
  \label{Fig:W-An-chi2}
\end{figure*}

Potentially significant background sources in this analysis are: 
$Z^{0} \rightarrow e^+e^-$; 
$W^{\pm} \rightarrow \tau\nu \rightarrow e\nu\nu$, 
where one of the final leptons is not detected; and events with an underlying 2-to-2 parton scattering (QCD events).
The first two sources have been evaluated using MC samples simulated with PYTHIA 6.4 using the ``Perugia 0'' tune. 
To estimate the relative contribution from background, the MC samples have been normalized to the $W^{+}$ and the $W^{-}$ data samples according to the collected luminosity.
In estimating the background from QCD events, we adopted the same ``data-driven'' technique used in previous STAR publications on $W^{\pm}$ production~\cite{STAR:Wlong,STAR:Wxsec}, reversing the 
selection criterion on $(\vec{P}_{T}^{bal} \cdot \vec{P}_{T}^{e}) / |\vec{P}_{T}^{e}|$ in order to select a data sample dominated by background. 
All background sources have been estimated to be at most a few percent of the selected sample, as reported in Table~\ref{Tab:BoS} and shown in Fig.~\ref{fig:plot_WZ_backgrounds}.

In the present work $A_N$ was also measured for $Z^{0}$ production, which is expected to be of the same magnitude~\cite{Kang:private} as for the $W^{\pm}$ boson and equally sensitive to the sign change of the Sivers function. The $Z^{0}$ bosons have a background with negligible impact to the spin asymmetry measurement and the kinematics are easily reconstructed from the two decay leptons produced within the acceptance of the STAR detector. Thus, the measurement is very clean and carries only the overall systematic uncertainty 
arising from the polarization measurement. 
The only experimental challenge is the much lower cross section of the $Z^{0}\rightarrow e^{+}e^{-}$ process, leading to poor statistics. 
The $Z^{0}\rightarrow e^{+}e^{-}$ events have been selected requiring two electrons with $P_{T} > 25$~GeV/c, of opposite charge, and with an invariant mass within $\pm 20 \%$ of the $Z^{0}$ mass value. Only 50 events 
remained after applying all the selection criteria. 

For each $y^{W}$ and $P_T^{W}$ bin, the data sample was divided into eight bins of the azimuthal angle $\phi$ of the produced boson, and the amplitude $A_N$ of the $cos(\phi)$ modulation was extracted fitting the following distribution,
which to first order cancels out false asymmetries due to geometry and spin-dependent luminosity differences~\cite{sqrtFormula}

\begin{eqnarray}
A_{N } &&   
= \frac{1}{\langle P \rangle}  \cdot  
\nonumber \\
&&\cdot 
\frac{\sqrt{N_\uparrow(\phi)N_\downarrow(\phi + \pi)} - \sqrt{N_\uparrow(\phi + \pi)N_\downarrow(\phi)} } 
{\sqrt{N_\uparrow(\phi)N_\downarrow(\phi + \pi)} + \sqrt{N_\uparrow(\phi + \pi)N_\downarrow(\phi)}},
\label{Eq:sqrtFormula}
\end{eqnarray}
where $N$ is the number of 
$W^+$, $W^-$, or $Z^0$  events
reconstructed
in collisions with an up/down ($\uparrow / \downarrow$) beam polarization orientation, and $\langle P \rangle$ is the average beam polarization magnitude.
Definining the up transverse spin direction $\vec{S}_{\perp}$ along the $y$-axis, and the direction of the polarized beam $\vec{p}_{beam}$ along the $z$-axis, the azimuthal angle is defined by $\vec{S}_{\perp} \cdot (\vec{P}^{W}_{T} \times \vec{p}_{beam}) = |\vec{P}^{W}_{T}| \cdot \cos(\phi)$.



The results for $A_{N}$ in $W^{+}$ and $W^{-}$ production are shown in Fig.~\ref{Fig:W-An} as a function of $P_{T}^{W}$ ({\it upper-left plot}) and the $W$ rapidity, $y^{W}$ ({\it bottom plots}). 
The absolute resolution in each of the three $y^{W}$-bins has been estimated to be $\sim 0.2-0.3$, whereas the relative resolution on $P_{T}^{W}$ decreases from $\sim 50\%$ in the first bin down to $\sim 30\%$ in the last bin.
The systematic uncertainties, shown separately by the shaded error bands in Fig.~\ref{Fig:W-An}, have been evaluated via MC. Events simulated by PYTHIA have been re-weighted as a function of $P_{T}^{W}$ and $y^{W}$ with the asymmetries as calculated by EIKV~\cite{Kang:2014} by comparing the generated and reconstructed distributions. The 3.4\% scale uncertainty on the beam polarization measurement is not shown in the plots.

For the $Z^{0}$ production, due to the low counts in the sample, $A_{N}$ was extracted for a single $y^{Z}$, $P_{T}^{Z}$ bin, following the same procedure used for the $W^{\pm}$, 
as shown in Fig.~\ref{Fig:W-An} ({\it upper-right plot}). 
The solid gray bands in Fig.~\ref{Fig:W-An} ({\it lower panels}) represent the uncertainty due to the unknown sea quark Sivers functions estimated by saturating the sea quark Sivers function to
their positivity limit in the KQ~\cite{Kang:2009bp} calculation. 

This analysis has yielded first measurements of a transverse spin asymmetry for weak boson production. 
The $A_{N}$ results as a function of $y^{W}$, shown in Fig.~\ref{Fig:W-An} ({\it bottom plots}), are compared with theory predictions from KQ, which does not account for TMD evolution, and from EIKV~\cite{Kang:2014}. The latter is an example among many TMD-evolved theoretical calculations (e.g., see~\cite{evol_papers}), though EIKV predicts the largest effects of TMD evolution among all current calculations.
Therefore, the hatched area in~Fig.~\ref{Fig:W-An} represents the current uncertainty in the theoretical predictions accounting for TMD evolution.
In contrast to DGLAP~\cite{DGLAP} evolution used for collinear parton distribution functions, TMD evolution contains, in addition to terms directly calculable from QCD, also non-perturbative terms, which need to be determined from fits to experimental data. A consensus on how to obtain and handle the non-perturbative input in the TMD evolution has not yet been reached~\cite{Collins}; therefore    
the results presented here can help to constrain theoretical models. 
A combined fit on $W^{+}$ and $W^{-}$ asymmetries,
$A_{N}(y^{W})$, to the theoretical prediction in the KQ model (no TMD evolution), shown in Fig.~\ref{Fig:W-An-chi2}, gives a $\chi^{2}/ndf=7.4/6$ assuming a sign-change in the Sivers function ({\it solid line}) and a  $\chi^{2}/ndf=19.6/6$ otherwise ({\it dashed line}).
The current data thus favor theoretical models that include a change of sign for the 
Sivers function relative to observations in SIDIS measurements, if TMD evolution effects are small.

We are grateful to Z.-B.~Kang for useful discussions.
We thank the RHIC Operations Group and RCF at BNL, the NERSC Center at LBNL, the KISTI Center in
Korea, and the Open Science Grid consortium for providing resources and support. This work was
supported in part by the Office of Nuclear Physics within the U.S. DOE Office of Science,
the U.S. NSF, the Ministry of Education and Science of the Russian Federation, NNSFC, CAS,
MoST and MoE of China, the National Research Foundation of Korea,
GA and MSMT of the Czech Republic, FIAS of Germany, DAE, DST, and UGC of India, the National
Science Centre of Poland, National Research Foundation, the Ministry of Science, Education and
Sports of the Republic of Croatia, and RosAtom of Russia.

\nocite{*}






\bibliography{apssamp}

\begin{thebibliography}{99}

\bibitem{Aybat:2011} S.~Mert~Aybat, and Ted~C.~Rogers, Phys. Rev. D {\bf 83}, 114042 (2011).

\bibitem{Meissner:2009} S.~Meissner, A.~Metz, and M.~Schlegel, J. High Energy Phys. 08 (2009) 056.

\bibitem{Sivers:1991} D.~W.~Sivers, Phys. Rev. D {\bf 41}, 83 (1990); D {\bf 43}, 261 (1991).

\bibitem{DIS_Sivers}
A.~Airapetian et al., the HERMES Collaboration, Phys. Rev. Lett. {\bf 94}, 012002 (2005); \\
M.~Alekseev et al., the COMPASS Collaboration, Phys. Lett. B { \bf 673}, 127 (2009); \\
X.~Qian et al., the Jefferson Lab Hall A Collaboration, Phys. Rev. Lett. {\bf 107}, 072003 (2011).

\bibitem{Collins:2002} J.~C.~Collins, Phys. Lett. B {\bf 536}, 43 (2002).

\bibitem{factorization}
S.~J.~Brodsky, D.~S.~Hwang, and I.~Schmidt, Phys. Lett. B {\bf 530}, 99 (2002); \\
S.~J.~Brodsky, D.~S.~Hwang, and I.~Schmidt, Nucl. Phys. B {\bf 642}, 344 (2002);\\
X.~Ji and F.~Yuan, Phys. Lett. B {\bf 543}, 66 (2002).

\bibitem{NSAC}
Nuclear Science Advisory Committee, the 2007 Long Range Plan, Milestone HP13 \\
\url{http://science.energy.gov/np/nsac/}. 

\bibitem{Kang:2014} M.~G.~Echevarria, A.~Idilbi, Z.-B.~Kang, I.~Vitev,
Phys. Rev. D {\bf 89},  074013 (2014).

\bibitem{E866}
E.~A.~Hawker {\it et al.}, Phys. Rev. Lett. {\bf 80},  3715 (1998).

\bibitem{Metz:2011}
  A.~Metz and J.~Zhou, Phys. Lett. B {\bf 700} 11 (2011).

\bibitem{Kang:2009bp} 
  Z.-B.~Kang and J.~-W.~Qiu, Phys. Rev. Lett.  {\bf 103}, 172001 (2009).
  
\bibitem{RHIC_polarimetry}
RHIC Polarimetry Group, RHIC/CAD Accelerator Physics Note 490 (2013). 

\bibitem{STAR_detector} K.~H.~Ackermann et al., the STAR Collaboration, Nucl.
Instrum. Methods Phys. Res., Sect. A {\bf 499}, 624 (2003).

\bibitem{TPC_detector} M. Anderson et al., the STAR Collaboration, Nucl. Instrum.
Methods Phys. Res., Sect. A {\bf 499}, 659 (2003).

\bibitem{BEMC_detector} M. Beddo et al., the STAR Collaboration, Nucl. Instrum.
Methods Phys. Res., Sect. A {\bf 499}, 725 (2003).

\bibitem{STAR:Wlong}
L.~Adamczyk {\it et al.},the STAR Collaboration, Phys. Rev. Lett. {\bf 113}, 072301 (2014); \\
M.~M.~Aggarwal {\it et al.},the STAR Collaboration, Phys. Rev. Lett. {\bf 106}, 062002 (2011).

\bibitem{STAR:Wxsec}
 L.~Adamczyk {\it et al.}, the STAR Collaboration, Phys. Rev. D {\bf 85}, 092010 (2012).

\bibitem{Tevatron-LHC-Wrecon}
D.~Acosta et al., the CDF Collaboration, Phys. Rev. D {\bf 70}, 032004 (2004); \\
G.~Aad et al., the ATLAS Collaboration, J. High Energy Phys. 12 (2010) 060;\\
S.~Chatrchyan et al., the CMS Collaboration, J. High Energy Phys. 10 (2011) 132.
	
\bibitem{PYTHIA6.4} T.~Sj{\"o}strand, S.~Mrenna and P.~Skands. J. High Energy Phys. 05 (2006) 026.

\bibitem{PerugiaTune} P.~Z.~Skands, Phys. Rev. D {\bf 82}, 074018 (2010).

\bibitem{RhicBOS}
P.~M.~Nadolsky and C.-P.~Yuan, Nucl. Phys. B {\bf 666}, 3 (2003);\\
P.~M.~Nadolsky and C.-P.~Yuan, Nucl. Phys. B {\bf 666}, 31 (2003).

\bibitem{GEANT3}
GEANT Detector description and simulation tool, CERN Program Library Long Write-up W5013, CERN
Geneva. 

\bibitem{Kang:private} Z.-B.~Kang, private communication.

\bibitem{sqrtFormula}
S.~B\"{u}ltmann {\it et al.}, Phys. Lett. B {\bf 632} 167 (2006);\\
S.~B\"{u}ltmann {\it et al.}, Phys. Lett. B {\bf 647} 98 (2007);\\
G.~G.~Ohlsen and P.~W.~Keaton Jr, Nucl. Instr. Meth. {\bf 109} 41 (1973).


\bibitem{evol_papers}
S.~M.~Aybat, A.~Prokudin, and T.~C.~Rogers, Phys. Rev. Lett. {\bf 108}, 242003 (2012);\\
M. Anselmino, M. Boglione, S. Melis, Phys. Rev. D {\bf 86}, 014028 (2012);\\
P.~Sun and F.~Yuan, Phys. Rev. D {\bf 88}, 114012 (2013).

\bibitem{DGLAP}
G.~Altarelli and G.~Parisi., Nucl. Phys. B {\bf 126}, 298 (1977);\\
Yu.~L.~Dokshitzer, Sov. Phys. JETP {\bf 46}, 641 (1977);\\
V.~N.~Gribov, L.~N.~Lipatov, Sov. J. Nucl. Phys. {\bf 15}, 438 (1972).

\bibitem{Collins}
J. Collins, EPJ Web of Conferences 85, 01002 (2015).

\end{thebibliography}

\end{document}